\documentclass[12pt,aps,preprint,showpacs,superscriptaddress]{revtex4}  %% REVTeX 4.0
\usepackage[draft]{hyperref} %% optional
\usepackage{graphicx}
\usepackage{setspace}
\usepackage{amsmath}
\usepackage{amssymb}
\usepackage[english]{babel}
\newcommand{\dcelcius}{$^{\circ{}}C$ }
\newcommand{\dcelciusn}{$^{\circ{}}C$}

\begin{document}

 \author{Jean-Michel Arbona, Jean-Pierre Aim\'e and Juan Elezgaray.}
\affiliation{CBMN, UMR 5248, CNRS, 2 r. R. Escarpit, 33600 Pessac, France.}
\title{Cooperativity in the annealing of DNA origamis.}

\begin{abstract}
DNA based nanostructures built on a long single stranded DNA scaffold, known as DNA origamis, offer the possibility
to organize various  molecules at the nanometer scale in one pot experiments.
 The folding of the scaffold is guaranteed by the presence of short, single stranded DNA
sequences (staples), that hold together separate regions of the scaffold. In this paper, we modelize the annealing-melting properties
of these DNA constructions. The model captures important features such as the hysteresis between melting and annealing, as well 
as the dependence upon the topology of the scaffold. We show that cooperativity between staples is critical to quantitatively explain the folding process
of DNA origamis.  

\end{abstract}

\maketitle

% DNA based nanostructures built on a long ssDNA scaffold, known as DNA origamis \cite{rothemund2006fdc}, are
% nowadays the basis of many applications, that range from single-molecule chemical reactions \cite{voigt2010single} to the organization
% at the nanometer scale of various molecules including proteins and carbon nanotubes \cite{maune2009self}. Curiously,
%  many basic questions about the mechanisms of formation of these constructions have not been
% addressed so far. For instance, the robustness of different designs against factors such as the internal 
% topology  are handled empirically. In this paper, we introduce and test a model for the folding and melting of DNA
% origamis. We show that several thermodynamic quantities measurable from UV absorption experiments can be reproduced accurately.
% The model can also be used to design a new distribution of crossovers that increases by almost 20 \dcelcius the melting temperature.

\section{Introduction}
DNA origamis \cite{rothemund2006fdc} are formed in an annealing process where a long single stranded DNA (ssDNA) scaffold 
extracted from a virus  (M13mp18)
is folded with the help of a set of $\sim 200$ short ssDNA strands, called staples. The resulting structure has a typical size of
100nm.  Locally, an origami is made of double stranded DNA (dsDNA) helices,
where one of the strands corresponds to the scaffold.
 Planar and 3D structures are obtained as the superposition of DNA helices, with basically only one constraint:
the staples should run from one helix to the other in such a way that mechanical deformations are minimized.
 This seemingly simple process has opened the way
to the fabrication of a variety of nanostructures in 2D or 3D, acting as biosensors \cite{KLC+08} or nanorabots 
\cite{lund2010molecular,gu2009dpp,sugiyama2011direct}. DNA origamis also provide with an easy tool to organize molecules
such as proteins or carbon nanotubes \cite{maune2009self} at the nanometer scale.

However, despite these innovative realisations, the folding process of DNA origamis remains poorly understood. 
Open questions include the exact role of the staple connectivity, the differences in yield for very similar origamis \cite{rothemund2006fdc},
or the factors that can influence the global stability of these nanostructures.
From an experimental point of view,
the process of formation of a DNA origami can be analysed by collecting the fluorescence intensity of a reporter dye \cite{castro2011primer}
or by monitoring the variation of its UV absorption as a function of the temperature (melting or annealing curve for, respectively,
increasing or decreasing temperatures) \cite{mergny2003analysis}. 
In this paper we study in depth the process of formation of DNA origamis through the analysis of melting and annealing curves
obtained from UV absorption experiments. 

The rest of the paper is organized as follows. Section 2 describes the formation of structures made of three ssDNA (called 'small origamis' in the following).
 Those can be considered as the 
building blocks of the usual DNA origamis. In this section, we also give 
a short description of the experimental techniques and the interpretation
of UV absorption data. In sections 3 and 4, the  model describing the energetic cost associated with the local connectivity of origamis,
and the modelling of the annealing-melting
processes are presented. Section 5 compares the melting curves of several DNA origamis with those predicted by the model.
The final section  presents some concluding remarks. A short report of this modelling \cite{oriepl} and a comparison between the
model and experimental AFM data \cite{oriAFM} have been published elsewhere.

\section{The basic folding mechanism unrevealed by small origamis.} 

\begin{figure}
\center
\includegraphics[width=0.45\textwidth]{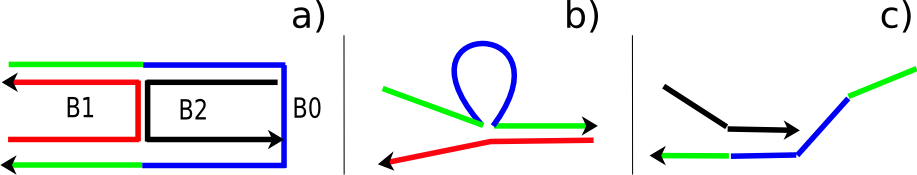}
\caption{(a) Schematic representation of the connectivity of the small origami.(b) and (c) Transitory situation correponding to the
binding of half of the staple to the scaffold. (b) B1 staple is in the 'outer' position,(c) B2 staple in the 'inner'
position. (b) and (c) show that the partial binding of staples in the outer (b) or inner (c) positions are very different.}
\label{fig:simpleShema}
\end{figure}

The basic mechanism in the formation of a DNA origami is the hybridization of a staple to the scaffold. In this section we will consider 
very simple systems that help in the understanding of this basic process. In order to simplify the reasoning, let us first consider the situation
where one staple (call it $S_1$) only binds to two separate parts of the scaffold (in Rothemund origamis, staples usually bind to three different parts of the
scaffold). The binding to the scaffold necessarily proceeds in two steps. First, one of the two parts of the staple
  binds to the correct region of the scaffold:
 the zipping mechanism that ensures this binding is stopped by the existence of a noncomplementary region in the scaffold. 
At this stage, half of the staple is bound to the scaffold, forming a stiff double helix. The remaining unbound parts of  the scaffold and the staple
form flexible coiled coil structures. Fig. \ref{fig:simpleShema} illustrates the two possible situations that can be found. When the staple is in the 'outer position',
the coiled coil structures are on the same side of the double helix. This is usually called a 'bulge' in the context of DNA thermodynamics 
\cite{santalucia2004thermodynamics}.
On the other hand, when the   staple is in the 'inner position', the coiled coil structures are on opposite sides of this double helix. Therefore,
the probability that they merge is lower and
the staple is expected to be much less stable in this position. In  Rothemund's origamis, both situations (inner and outer) are found. Whereas
the formation of a bulge (outer position) is well documented, it seems that the thermodynamics of the 'inner position' binding has not been considered previously.
 This is why we decided to study
the small origami structures.

Another important aspect in the folding of DNA origamis is the interaction between staples. Following the simple reasoning of the previous paragraph,
let us consider the influence of a second staple $S_2$ on the binding of $S_1$. If the melting temperature of $S_2$ is higher than that
of $S_1$ ($T_m(S_2) > T_m(S_1)$), by the time $S_1$ starts binding, with high probability both parts of $S_2$ are bound to the scaffold, reducing its effective elasticity and/or 
length. This could in turn facilitate the binding of the second part of $S_1$, resulting in some kind of cooperative behaviour.

We designed a DNA construction (called small origami) made of two ssDNA 32b long (staples) and a 64b long ssDNA (scaffold).
This structure is similar to DAO structures\cite{fu1993ddc} and comparable in size and shape to JX and PX structures which have already been studied experimentally \cite{spink2009thermodynamics},
and theoretically \cite {maiti2004stability}, \cite{maiti2006atomic}. 
Three different sets of staples based on the same structure were chosen to quantitatively evidence cooperative effects during the binding of the staples.
In the first two sets, the two staples (B1 and B2) have very different compositions: the sequence of B1 only contains A or T nucleotide whereas 
B2 only contains G or C nucleotide. Accordingly, their melting temperatures are far apart, respectively 57 \dcelcius and 91 \dcelcius.  
This allows to  differentiate the two staples in the melting curve.
The third set has two staples, B1m and B2m, designed with chemical sequences different enough to avoid mispairing with the 64b template B0. They have close melting temperatures (respectively 77 \dcelcius and 80 \dcelcius) as their AT/GC ratio are similar.

The topology of the binding is illustrated in (Fig.~\ref{fig:simpleShema}): each staple contains two contiguous parts, 32b long,
that bind to the scaffold. In (Fig.~\ref{fig:simpleShema}a), B1 is in the 'outer' position, B2 is in the 'inner' position. 
%We have already explained the difference between these two ways of binding. 
%can be further stressed by considering what happens when only half of the staple is hybridized.
%In the 'outer' position, the unbound parts of the staple and the scaffold are located on the same side of the bound moities 
%(Fig.~\ref{fig:simpleShema}b). In the 'inner' position, the unbound parts are on opposite sides (Fig.~\ref{fig:simpleShema}c). 
It should be noted that  besides the existence of an entropic hindrance, 
the inner position requires that double-helical domains stay in close contact, which could result in additional instability.
 Motivated by the previous considerations, we have investigated three different cases:

\begin{itemize}
\item  B2  is located in the inner part, B1 in the outer part.
\item  B2  is located in the outer part, B1 in the inner part.
\item B1m is located in the outer part, B2m in the inner part.
\end{itemize}

For each of these three cases, we have performed 
UV absorption measurements as a function of temperature. These measurements are based on the fact that hybridized bases absorb less than open bases: this is the so-called hyperchromic effect.
The absorbance associated to any staple $S_i$ is:

\begin{equation} \label{absorption}
Abs_{S_i}(T) = \left(Abs_0(S_i) + Abs_0(\bar{S_i})\right)\left( 1-h_{S_i} x_{S_i}(T)\right). 
\end{equation}
In this equation, $x_{S_i}(T)$ denotes the proportion of staples that are folded (at temperature $T$),
 $Abs_0(S_i)$ is the absorbance of  $S_i$ and $Abs_0(\bar{S_i})$ the absorbance of its complementary staple, calculated according to \cite{tataurov2008predicting}. We used the following relation for the hypocromicity $h_{S_i}$:

\begin{equation}
h_{S_i}= 0.287 (1-f_{gc}(S_i)) + 0.15 f_{gc}(S_i)
\end{equation}
where $f_{gc}(S_i)$ is the fraction of GC content in the staple $S_i$.
This formula has been slightly modified from \cite{tataurov2008predicting} to improve the fit against the experimental
data for the small origamis. The degree of pairing $\theta(T)$ can be obtained from the raw absorbance measurements
following the methods in \cite{mergny2003analysis}.

\begin{figure*}
\center
\includegraphics[width=1.\textwidth,height=0.23\textheight]{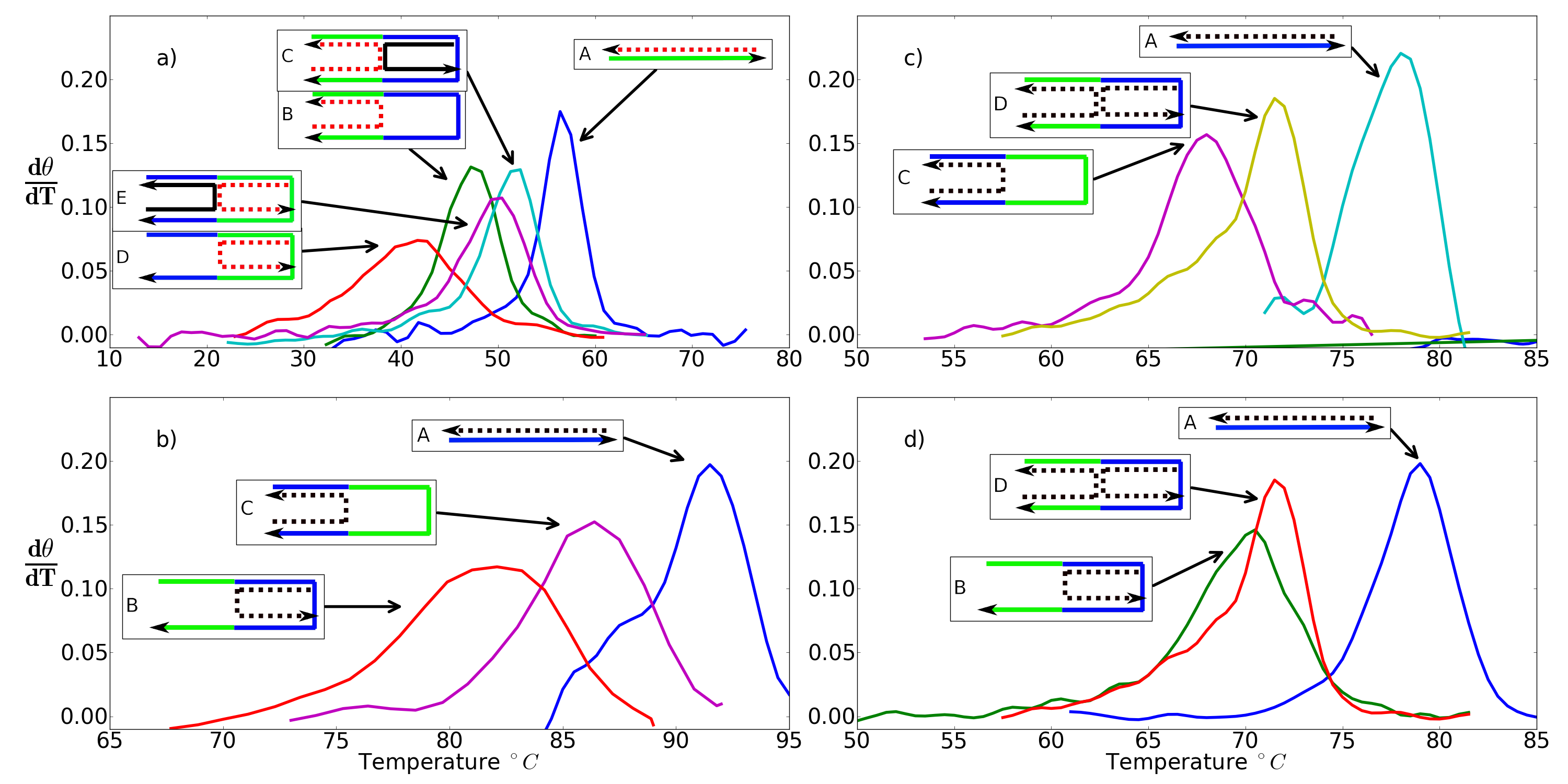}

\caption{ The derivative $d\theta/dT$ reported in the four figures corresponds to the folding of the dotted staple. 
a) experimental data on the folding of B1(AT) cases (A,B,D) in the absence of B2(GC), cases (C,E) with B2 already folded;
b) experimental data on B2 without B1; 
c) experimental data on B1m; 
d) experimental data on B2m }
\label{fig:allsmall}
\end{figure*}

In (Fig.~\ref{fig:allsmall}a) we report the derivative of the melting curves $\theta(T)$ that show the behaviour of the B1 strand for different configurations.
%:B1 with its complementary strand B1*, B1 alone in the outer position, B1 in the outer position with B2, B1 alone in the inner position,
%and B1 in the inner position with B2. 
Fig.~\ref{fig:allsmall}a.A corresponds to the melting curve of  B1 with its complementary $\overline{B1}$: it shows a maximum peak at 57~\dcelcius and a half width of 4.5 K. 
We analyse first the case where B1 is outer. Without the staple B2 (Fig.~\ref{fig:allsmall}a.B),  the structure produces a loop or bulge, as described in \cite{santalucia2004thermodynamics}
that introduces an entropic penalty and decreases the melting temperature to 48.5~\dcelcius and a half width of 6 K. 
Thus the folding is much less robust. When B1 and B2 are both present two events appear on the melting curve. 
The first one at 83 \dcelcius  (not shown in Fig.~\ref{fig:allsmall}a) corresponding to the folding of B2 in the inner position.
Then Fig.~\ref{fig:allsmall}a.C,  B1 folds at a temperature higher than when it is alone, with a maximum peak at 51.5~\dcelcius but with a similar half width. Therefore, the inner staple B2 helps the pairing of the outer staple B1  by suppressing part of the entropic penalty related to the bulge formed by the scaffold. 
When B1 is in the inner part (Fig.~\ref{fig:allsmall}a.D), its binding is significantly destabilized: the maximum peak is located at 41.5 \dcelcius, 15 K lower than the value of the dsDNA and with a half width as large as 12 K. 
This is three times larger than the one of B1$\overline{B1}$. Again, when the B2 staple is added  (Fig.~\ref{fig:allsmall}a.E ), the pairing of B1 is stabilized with a maximum peak much higher located at 50 \dcelcius but still with a rather large half width of 8 K. 
These experimental results support the evidence of a strong correlation between the two strands: 
the presence of B2 helps the folding of B1 whatever its location. Moreover, these results also show that the location of the strand is of importance,
 the inner location being much less favourable.
% The origin of this difference is not obvious, it may in part be the result 
%of an entropic penalty larger than the one a bulge induces, as it occurs when B1 is located in the outer domain, but it may also be the consequence 
%of the energy cost of a local curvature the inner strand imposes to the B0 template.

Because B1 folds at a much lower temperature than B2, the binding of B2 is not influenced by the presence of B1.
Fig.~\ref{fig:allsmall}b illustrates the influence of the location (inner or outer) for B2. The double helix structure 
B2$\overline{B2}$ (Fig.~\ref{fig:allsmall}b.A) has a maximum peak at 91\dcelcius with a half width of 3.5 K, while when B2 is in the inner position (Fig.~\ref{fig:allsmall}b.B) the peak is located at 83 \dcelcius 
with a half width of 8 K. When B2 is in the outer position (Fig.~\ref{fig:allsmall}b.C) the peak is located at 86.5\dcelcius and is
 slightly narrower (half width of 6.5 K). 
Therefore, the same differences between the inner and outer positions are observed whatever the chemical sequence involved.

The experimental results obtained with the B1m-B2m set of staples are shown in (Fig.~\ref{fig:allsmall}c and  \ref{fig:allsmall}d). The same trends as
in the B1-B2 case are observed but with less pronounced effects:
 the temperature shifts and the increase of the half width  with respect to 
the melting curves of the double strands B1m$\overline{B1m}$ and B2m$\overline{B2m}$ are smaller. A correlation effect is also noticeable, and is now observed for both strands B1m and B2m. 
When the staple B1m is in the solution, the staple B2m, which folds at a higher temperature, shows a narrower peak at a maximum 1.5 \dcelcius higher than when it is alone. 
Therefore partial folding of the staple B1m helps the folding of the staple B2m. Similarly to what was observed in the previous case, the correlation effect is even 
more effective when we consider the influence of the B2m staple on the folding process of the B1m staple (Fig.~\ref{fig:allsmall}c.D)
with a shift of the melting temperature from 67 \dcelcius to 72 \dcelciusn.

\section{Hypothesis for the model of DNA origami folding.}
\label{sec-1}

The previous section considered the folding of simple DNA constructions that go beyond the simplest
double stranded DNA structure. A visual inspection of Rothemund DNA origamis shows that small origamis can
be considered as their building blocks. Therefore, we expect that the same physical mechanisms (cooperativity and
topological effects) that explain the formation of small origamis will be present
in the formation of larger origamis. However, we first need to make a few hypothesis in order to make the problem tractable.
The basic difficulty is related to the huge number of possible configurations that need to be handled 
to compute average properties such as the number of open base pairs. Long linear structures of double stranded
DNA can be computed rigorously because recurrence relations can be established in such cases \cite{poland1970theory},\cite{jost2009unified}.
DNA origamis are highly connected structures bearing pseudoknots. This prevents the use of linear recurrences. 

The DNA origami folding process can be generally  described by a set of ’insertion’ reactions of the form
\begin{equation} S_i + N(S_i)\rightleftharpoons N(S_i)S_i, \label{eqins} \end{equation}
where $S_i$ denotes the $i-$th staple, and $N(S_i)$ represents a particular set of neighbour staples
(to be defined below) that influence the insertion of staple $S_i$. Eq. \ref{eqins} will be made more precise in the following.
To proceed further, we need to make further approximations.

Each staple
$S_i$ of length $|S_i|$ can be divided in parts that hybridize to non-contiguous regions of the scaffold. Let us note
$S_i = S_{i,1} + S_{i,2} + \ldots $ such a division of the strand sequence (typically,
each 32b staple is divided in three parts but other partitions are possible). Staples in the
small origamis are divided in two parts: $S_i = S_{i,1} + S_{i,2}$.

{\bf  Hypothesis 1}: we will focus  on configurations where each part $S_{i,k}$ is either completely hybridized to the
scaffold or completely 
unbound. Moreover, we also disregard misfolded configurations, that is, staples that partially hybridize to the 'wrong'
part of the scaffold. Notice that this assumption is plausible for the one layer origamis we consider here. For
more complex, multi-layered structures, the staples are divided in smaller parts so that the probability to bind
to the wrong part of the scaffold is considerably increased. We also disregard possible secondary structures
of the scaffold.

%\subsection{Calculation of the different possible states (Step2)}

We will call {\em crossover} the connection between two contiguous parts of any staple. A crossover is not
associated with a particular DNA base, it is only a convenient notation to describe the connectivity of the
origami. In Rothemund's origamis \cite{rothemund2006fdc}, typical staples are 32 bases long and composed of three parts (8,16,8 bases long respectively) linked by two crossovers $cS_{i,1}$ and $cS_{i,2}$.
On the scaffold side, a crossover is associated with a loop, a subset of the scaffold that is
hybridized (or not) depending on the presence of  other staples.  In the previous example (Fig.~\ref{fig:simpleShema}a) B1 and B2 are composed of
 two parts connected  by a crossover,  and B0 plays the role of the scaffold.

{\bf Hypothesis 2}: configurations with non contiguous hybridized parts are forbidden. This
hypothesis is well verified when the central part of the staple is much longer than the other parts. In the following,
we will note $S_i(k,l) (k \le l)$ the configuration where $S_{i,k}$, $S_{i,k+1}, \ldots $, 
$S_{i,l}$  are hybridized, the other parts being unpaired. 

The model aims to compute the probability $p(S_i(k,l),T)$ of having a particular folded state of the staple
$S_i$ at temperature $T$. We will assume that at very high temperature $T=T_h$, all the  staples are unfolded: 
$p(S_i(k,l),T_h)=0 $ (in practice, $T_h=90$\dcelcius).
 The model is {\em recursive}: $p(S_i(k,l),T+dT)$ is computed based on the knowledge of $p(S_i(k,l),T)$.
The increment $dT$ can be  positive or negative: the algorithm starts from $T_h$, the temperature 
decreases down to a value $T_{l}$, then increases again. At any temperature $T$, the probability to observe
  a given configuration $S_i(k,l)$ will depend upon the presence (or not) of neighbour staples. Therefore, for each staple
a set of possible neighbour staples $\{ N_1(S_i), N_2(S_i), \ldots \}$ needs to be defined. In the following, we will
use the generic notation $N_{\alpha}(S_i)$ for any of these neighborhoods.

How many staples one has to consider in each of these sets
is a parameter of the model. In this paper, all the origamis are such that the crossovers can be 
aligned in rows. In the following, for each staple and crossover, 
the set of neighbour staples will be limited to those that have crossovers in the same row, and
 are separated by less than 75b. Let us note that all the results on this paper are robust against variations
of this parameter. With
these notations, the probability to observe the staple $S_i$ in a given configuration $ S_i(k,l)$ and for
a given neighbourhood $N_{j}(S_i)$ is modelized by an equilibrium reaction:
\begin{equation} S_i(k,l) + N_{\alpha}(S_i)\rightleftharpoons N_{\alpha}(S_i)S_i(k,l). \label{eqins1} \end{equation}
This modelling therefore does not consider any kinetic effect. Notice however that the equilibrium hypothesis
is a {\em local} one. Indeed, we expect that the model will reflect the hysteresis observed in the annealing-melting
of DNA origamis.

{\bf Hypothesis 3}: because the model only keeps track of the single 
probabilities $p(S_i(k,l),T)$ and not of the joint probabilities $p(S_1(k,l),S_2(k',l'), \ldots T)$, it is
necessary to make an additional approximation to determine $p(N_{\alpha}(S_i),T)$. Again, based on the data
from  the small origamis, we assume that there is a strong correlation between the different staples.
%At a given temperature $T$, all of the staples $S_i$ with $T_m(S_i) > T$ can be considered as hybridized.
As the processes of annealing and melting are monotonous, for two staples $S_{i_1}$ and 
$S_{i_2}$  we make the hypothesis that if $p(S_{i_1}(k,l),T)<p(S_{i_2}(k',l'),T)$, then the $S_{i_2}$ staple
is present in the structure when $S_{i_1}$ starts to fold.
  In order to compute $p(N_{\alpha}(S_i),T)$, let us generalize this idea and order the staples
in $N_{\alpha}(S_i)= \{ S_{i_1}, S_{i_2}, \ldots \}$ in such a way that $p(S_{i_1},T) \leq  p(S_{i_2},T) \leq \ldots < p(S_c)=1,$
where $S_c$ stands for the scaffold.
According to the {\em high correlation hypothesis}, we approximate the joint probabilities in the following
way:
\begin{eqnarray*}
p(S_{i_1},T) & = & p(S_{i_1},S_{i_2}, \ldots ) \\
p(S_{i_2},T)-p(S_{i_1},T) & = & p(S_{i_2},S_{i_3}, \ldots ), \ldots
\end{eqnarray*}

For instance, in the case where only two crossovers influence $S_i$, with probability
$p(S_{i_1})$ both $cS_{i_1}$ and $cS_{i_2}$ are present, the probability of only having $cS_{i_2}$
is $p(S_{i_2})-p(S_{i_1})$ and the probability of only having the scaffold is $1-p(S_{i_2})$. 

The probability $p(S_i,T)$ for staple $S_i$ to be hybridized at temperature $T$ is the solution of
the coupled set of equations:
\begin{equation}
p(S_i,T) = \sum_{N_{\alpha}(S_i)} p(S_i,T | N_{\alpha}(S_i)) p(N_{\alpha}(S_i),T). \label{eqcl}
\end{equation}
This set of equations can be further approximated by the iteration: 
  \begin{equation}
p(S_i,T) = \sum_{N_{\alpha}(S_i)} p(S_i,T | N_{\alpha}(S_i)) p(N_{\alpha}(S_i),T-dT). \label{eqcl_app}
\end{equation}
Here, $dT$ denotes the temperature step that determines the annealing-melting protocol
and $p(S_i,T | N_{\alpha}(S_i))$ the conditional probability to hybridize staple $S_i$ in the
neighborhood $N_{\alpha}(S_i)$.
In the annealing process ($dT < 0$), 
at high temperature $T_{\infty}$, $p(N_{\alpha}(S_i, T = T_{\infty}))=0$ 
excepted for the empty neighborhood (the staple hybridizes to the scaffold in the absence of
any other staple).  For lower temperatures, $p(N_{\alpha}(S_i),T)$ can be computed
from the knowledge of $p(S_i,T)$ and the correlations between staples (hypothesis 3).
We show in appendix A that the set of coupled equations \ref{eqcl} determines  $p(S_i(k,l),T)$ 
provided the equilibrium constants of the reactions \ref{eqins1} are known. This amounts
to define an energy model which is detailed in the next section.

\section{Modelling the energy of DNA origamis}
\label{sec-1_2}

In this section, we introduce an energy model for the hybridization of staples to the scaffold of a DNA origami.
The gain in Gibbs free energy for hybridizing $S_i(k,l)$, 
$ \Delta G (S_i(k,l),T) =  \Delta H (S_i(k,l),T) -  T\Delta S (S_i(k,l),T)$
contains two contributions: 
\begin{equation} \Delta G = \Delta G_{NN} + \Delta G_{top}.\end{equation}
 The local contribution
$\Delta G_{NN}$ only depends on the sequence of $S_i(k,l)$. It quantifies the gain in free energy associated with
the local formation of  double helices. We use the parameters of the 
nearest-neighbour model \cite{santalucia2004thermodynamics} 
with a temperature correction given in \cite{hughesman2011correcting} (see appendix \ref{TScorr} and 
\cite{hughesman2011correcting} for a detailed 
description). 

\begin{figure}
\center
\includegraphics[width=0.3\textwidth]{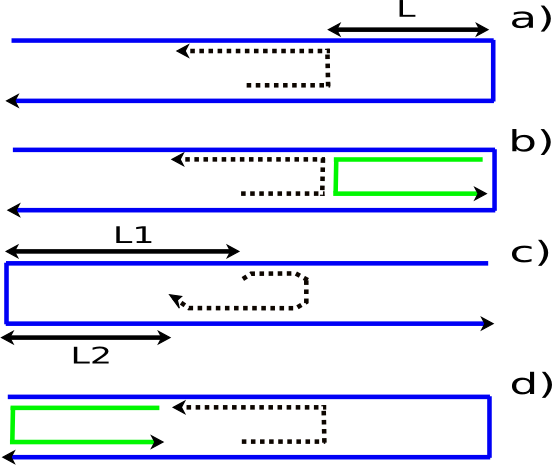}
\caption{representation of the three different local intermediate states (LIS). The staple to be inserted is
represented by the dotted line, the scaffold by the continuous line.
(a) LIS1 corresponds to the formation of a bulge. (b) LIS2 corresponds to the case where, due to the binding of 
other staples, the effective length of the bulge is zero. (c) LIS3 generalizes the 'inner position' of the
small origamis. We assume that, because of the
curvature constraints imposed by this configuration, the staple is not able  to fold completely. (d) A typical situation where
two types of LIS (LIS1 at the right side of the staple, LIS3 at the left side) 
can be attributed to a given crossover.}
\label{fig:twodefect}
\end{figure}

$\Delta G_{top}$ gathers several contributions that depend on the  connectivity of the origami ($\Delta G_{NN}$ depends
mostly on the sequence of the scaffold and, to a less extent, on the density of crossovers, but not on the 
connectivity). With any crossover, we associate  an entropic penalty. This penalty reflects the
difficulty for a staple to hybridize  non-contiguous parts of the scaffold.
 In a first approximation, the longer the region of the scaffold that connects
the two parts to be hybridized, the larger the penalty. Our previous results obtained  with the small 
origami show that this needs to be refined. Based on these data, we  consider three  situations
characterized by transient arrangements of staples that we call {\em local intermediate states} (LIS).
 In the first one (LIS1), the staple hybridizes
to the scaffold, forming an internal asymmetric loop \cite{santalucia2004thermodynamics}, Fig.~\ref{fig:twodefect}a.
 The length of this loop corresponds to the number of unpaired bases of the
scaffold linked by the crossover. This is a generalization of the 'outer' position found for
 the three strands origami.
Before the crossover formation, when only  part of the staple is folded,
the scaffold and the non hybridized part of the staple  are on the same side of the hybridized part of the staple (Fig.~\ref{fig:simpleShema}b).
 In this case, the staple is not involved in the path that connects
the two extremities of the crossover.
A particular case of LIS1, which we call LIS2, is the situation where the 
length of the loop is zero: the  crossover forms locally a Holliday junction, the staple hybridizes in the close
vicinity of an already hybridized staple (Fig.~\ref{fig:twodefect}b). 

The third LIS, LIS3, corresponds to the inner position in the small origami: the shortest path that connects the two ends
of the crossover involves the staple itself (Fig.~\ref{fig:twodefect}c). Because of this, before the crossover forms, the non hybridized
parts of the strand and  the scaffold are located on opposite sides of  the hybridized parts (Fig.~\ref{fig:simpleShema}c).
Therefore,  LIS3 implies a larger penalty than LIS1 or LIS2. In the small origami, the shift in $T_m$ was less than 10 \dcelcius for LIS1, between 
5 \dcelcius and 7 \dcelcius for LIS2 and up to 15 \dcelcius for LIS3.

\begin{figure}
\center
\includegraphics[width=\textwidth]{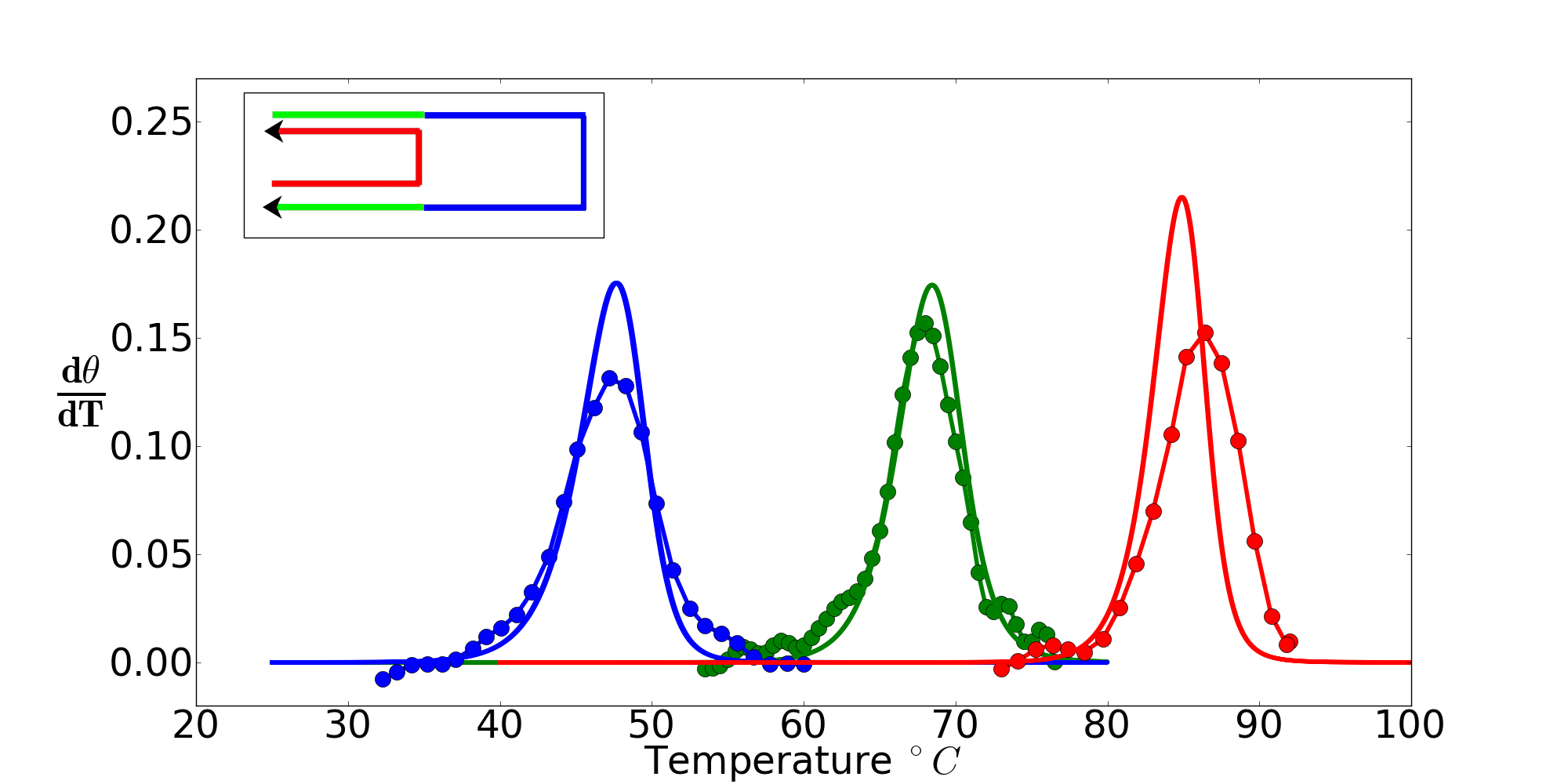}
\caption{Melting curves obtained for a small origami containing only one staple in the outer position (inset).  
Comparison between the experimental (dotted line) derivative $d\theta/dT$ and the modeling (continuous line) as provided
by the contribution $\Delta G_{top}$(LIS1). In order of increasing melting temperature: B1 staple (blue), B1m staple (green)  and B2 staple (red).}
\label{fig:brinloopdefect}
\end{figure}

\begin{figure}
\center
\includegraphics[width=\textwidth]{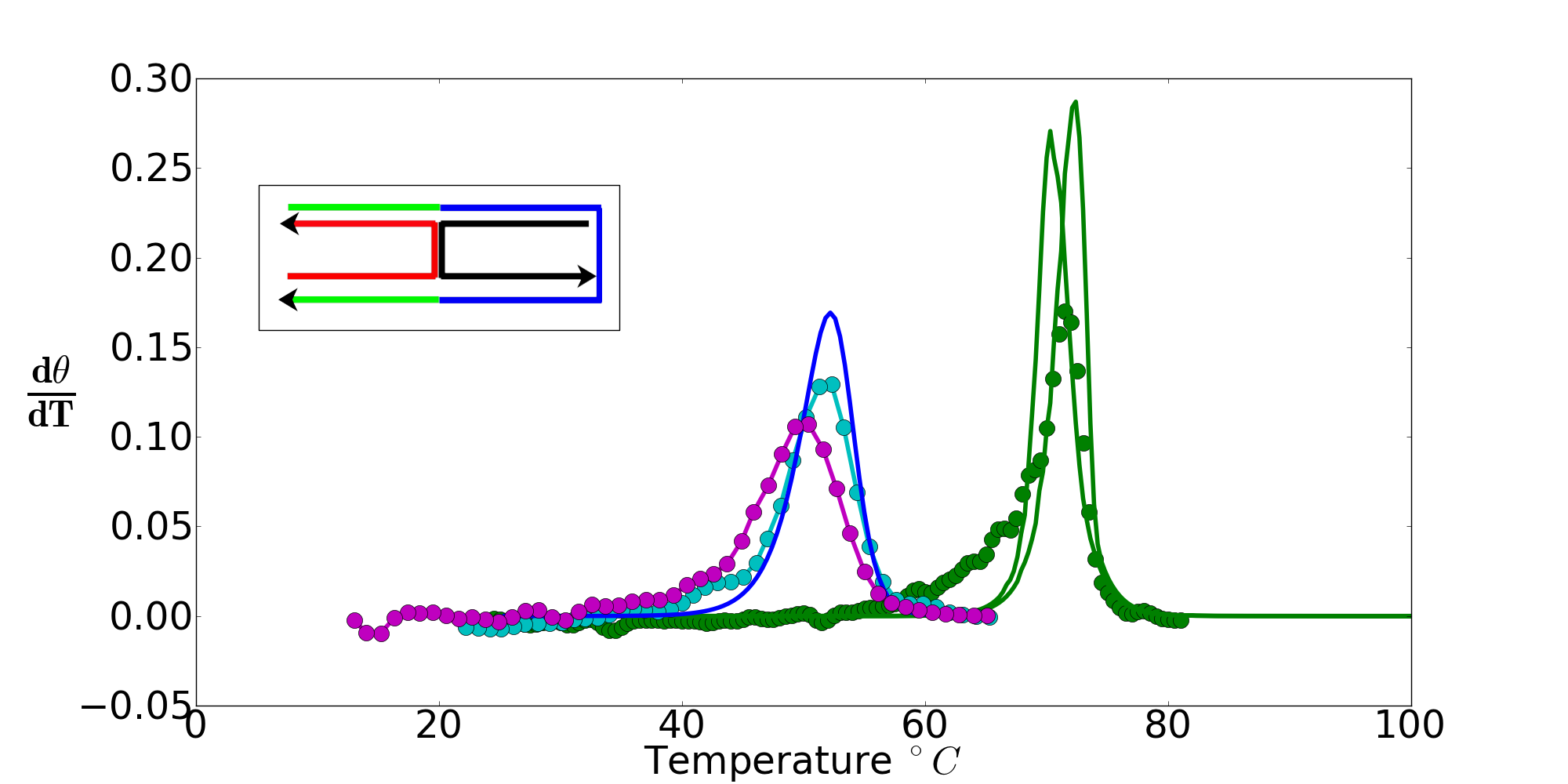}
\caption{Melting curves obtained for a small origami containing  two staples  (configuration represented in the inset).  
Comparison between the experimental (dotted line) derivative $d\theta/dT$ and the modeling (continuous line) as provided
by the contribution $\Delta G_{top}$(LIS2). The curves clustered around $T_m \sim 50$ \dcelcius correspond to B1 in position outer (cyan),B1 in position inner (magenta), the modelling does not make any difference between these two cases. The curves clustered around $T_m \sim 70$ \dcelcius
correspond to the simultaneous binding of B1m (inner) and B2m (outer).}
\label{fig:brinscouple}
\end{figure}

\begin{figure}
\center
\includegraphics[width=\textwidth]{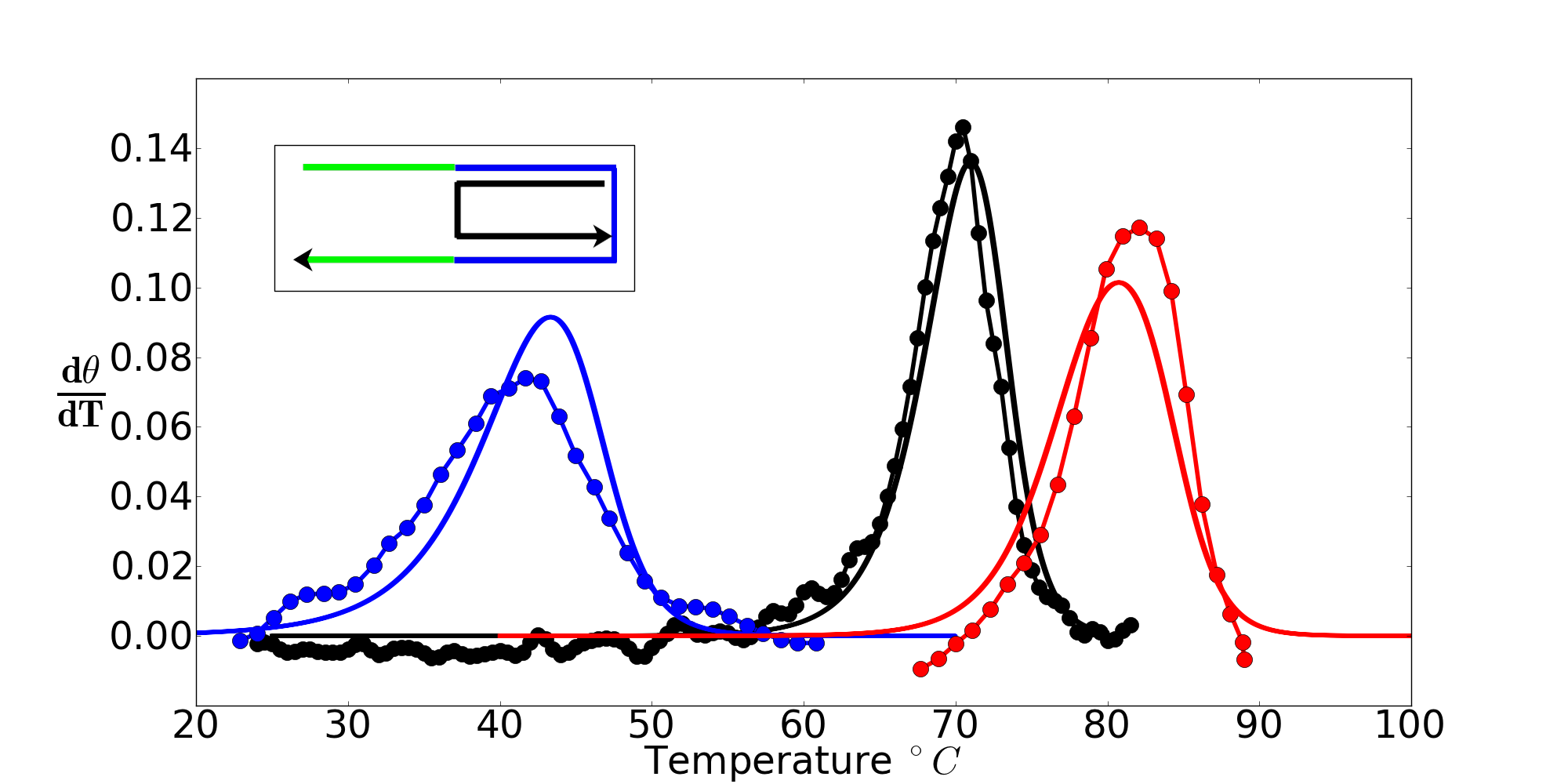}
\caption{ Melting curves obtained for a small origami containing only one staple in the inner position (inset).  
Comparison between the experimental (dotted line) derivative $d\theta/dT$ and the modeling (continuous line) as provided
by the contribution $\Delta G_{top}$(LIS3). In order of increasing melting temperature:  B1 staple (blue), B2m staple (black),B2 staple (red).}
\label{fig:brinsinerc} 
\end{figure}

To each of these LIS is associated a different $\Delta G_{top}$:
\begin{itemize}
\item  $\Delta G_{top} $(LIS1)$= -T\Delta S_{bulge}(n_T - 0.8<nb_{folded}>)$. The function $\Delta S_{bulge}(n_T)$ is that of ref.
 \cite{santalucia2004thermodynamics}. $n_T$ corresponds to the number of bases along the scaffold and $<nb_{folded}>$ is the
average value of bases folded along the scaffold (this average takes into account the probabilities of all the possible neighbouring configurations).
The comparison between this model and the experimental results from the small origami structure is illustrated in Fig.~\ref{fig:brinloopdefect}. 

\item $\Delta G_{top} $(LIS2) $ = \Delta H_{loop0}-T \Delta S_{loop0}$ with $\Delta H_{loop0} = 25.3  kcal/mol$ and $\Delta S_{loop0} =65.0  cal/mol/K$. This constant contribution has been derived so as to fit as well as possible the B1-B2 experimental data (left of Fig.~\ref{fig:brinscouple})
 and then applied to the B1m-B2m data (right part of Fig.~\ref{fig:brinscouple}).

 At the crossover,  two bases that belong to  $S_i$  face the bases constituting the crossover made by the other strand. 
The initial enthalpic and entropic contribution of this pair of bases  is subtracted from $\Delta G_{NN}$ as they are not nearest-neighbours
anymore. One half of the contribution (nearest-neighbour model) of the two new pair of bases is added.
\item 
$\Delta G_{top} $(LIS3) $= \Delta H_{unbind}-T\Delta S_{unbind} -T\Delta S_{bulge}(n_t-0.8<nb_{folded}>)$ (Fig.~\ref{fig:twodefect}c): an additional penalty is added to 
the entropic penalty of the loop. This intends to reflect the stronger instability characteristic of this LIS. $\Delta H_{unbind}$  (resp $\Delta S_{unbind}$)
quantifies the loss of enthalpy (resp entropy) associated with the partial unfolding of the ends of an staple involved in this type of LIS. The number
of bases that unfold is a parameter of the model. The data in  Fig.~\ref{fig:brinsinerc} correspond to the unfolding of a total of 8 bases
(two bases for each of the four extremities of the staple, see Fig.~\ref{fig:twodefect}c).
\end{itemize}

Under some circumstances (Fig.~\ref{fig:twodefect}d), two types of LIS can be attributed to a given crossover.
In such cases, the LIS with the smaller $\Delta G_{top}$ is taken into account.

The modelling obtained with the contributions $\Delta G_{NN}$ and $\Delta G_{top}$ is quite satisfactory except for a constant
negative shift ($\sim -4$K) of the melting temperatures. This shift indicates that another stabilizing mechanism  that is not present in 
small constructions such as the small origamis has to be invoked. Indeed, in the folded structure of origamis, double-helix sections are
separated by distances of the order of 1nm. It is then reasonable to think that mechanisms such as correlations between counter-ions and
hydration forces come also into play, as is the case when DNA condensates \cite{SPR+98}.
This electrostatic stabilizing term
 is proportional to the number of neighbouring bases $n_{n} $ that are close to the staple $S_i$, and to the 
length $|S_i(k,l)|$ of the partial configuration  $S_i(k,l)$ considered. 
\begin{equation}
\Delta H_{n} (S_i(k,l))= -0.132 n_{n} |S_i(k,l)|/|S_i| (kcal/mol)
\end{equation}
The energy per base  $-0.132 kcal/mol/base$ is similar to the one needed for DNA condensation $10^{-1} k_B T / base$ to $10^{-2} k_B T / base$ \cite{bloomfield1997dna}. An alternative explanation for the presence of this term 
would be the entropy reduction due to the confinement of the double helices in a 2D structure. 

The probability to fold any part $S_{i,j}$ of staple $S_i$ is given by the sum of the probabilites of any configuration
$S_i(k,l)$ that contains $S_{i,j}$:
\begin{equation}
p(S_{i,j},T) = \displaystyle{\sum_{l,p, l \le j \le p} p(S_i(l,p),T).}
\end{equation}
The fraction of folded bases for staple $S_i$ 
\begin{equation}
x_{S_i}(T) = \frac{1}{|S_i|}\sum_j p(S_{i,j},T)|S_{i,j}|
\end{equation}
can be then  converted to a theoretical absorbance \cite{tataurov2008predicting} as in the case of small origamis \ref{absorption}.

\section{Comparison with the annealing-melting data of DNA origamis}
\label{sec-2}

%figure_poster

\begin{figure*}
\center
\includegraphics[width=.9\textwidth,height=0.23\textheight]{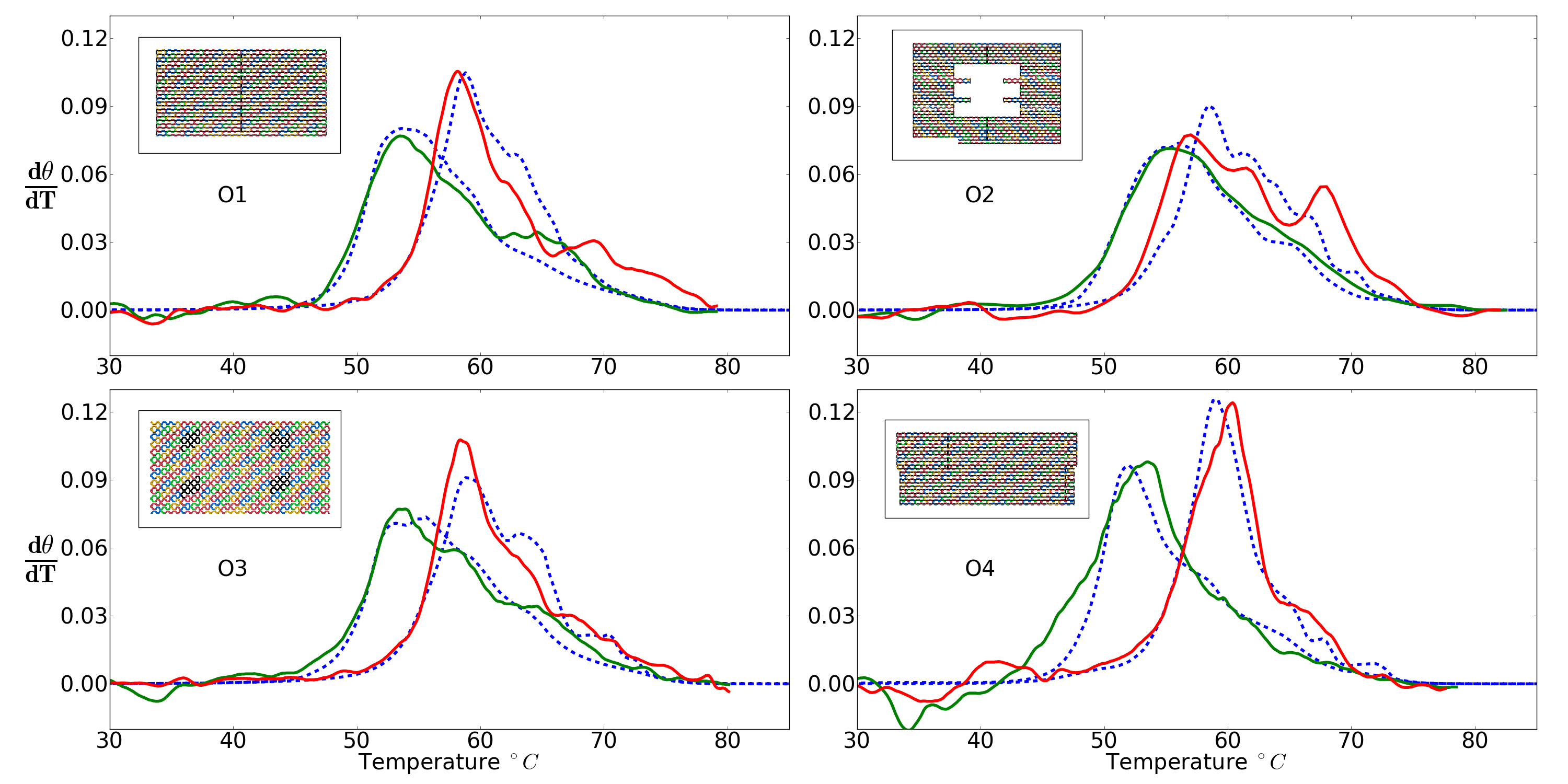}
\caption{Derivative of the pairing degree versus temperature for the O1, O2, O3 and O4 origamis
represented in the insets. The data corresponding to annealing are in red, 
melting  in green, the model (dashed line) is in blue
for both processes.}
\label{fig:drawing}
\end{figure*}

We considered four different DNA origamis (Fig.~\ref{fig:drawing})
with the same scaffold (M13mp18 virus) and about 200 staples. O1 is the rectangle in the original Rothemund work \cite{rothemund2006fdc}, the staples
are mostly 32b long, with folding part sequences  divided in 8-16-8 patterns. 
O2 is another rectangular origami that includes a hole \cite{endo2010regulation} and
presents the same 8-16-8 pattern. O3 has the same connectivity pattern as O1, but some staples have been merged two by two in four areas (coloured in black in \ref{fig:drawing}c),
so that the typical staple pattern is 8-16-16-16-8. Finally, O4 is another rectangular origami where a 100b long subset of the scaffold goes
from one side to the other of the rectangle, forming a ssDNA 'bridge'.
 
For each origami, a series of annealing-melting cycles was performed coupled to UV-absorption measurements. With the methods
of \cite{mergny2003analysis}, the degree of pairing $\theta(T)$ was extracted as a function of temperature. The temperature ramp 
(0.4 \dcelcius min$^{-1}$) is typical for this one-layer origamis. The annealing-melting process is not symmetrical, the hysteresis
between the two phases of a cycle is such that the melting takes place at temperatures higher than the annealing.

There is an overall agreement (Fig.~\ref{fig:drawing}) between the melting-annealing curves observed experimentally and computed with the model.
The model captures the hysteresis between the annealing and melting processes, as well as the relative strength of this hysteresis between different
origamis (O2 has only 4K shift between annealing-melting, whereas O4 has 10K shift). The maximum value of the derivative, which
can be linked to the overall enthalpy of the transition in a two-state model, is also reproduced. This feature is  robust against
small variations of the parameters of the model.
  
%The main differences are reproduced:  the very low hysteresis for the Rectangle h, the high of the derivatives for both are very different
%and reproduced. For the Rectangle (Fig.~\ref{fig:drawing}) there is a miss for the theoretical derivatives at high temperature.
%In ordered to understand the differences between those three origamis, we decided to change different parameter of the design
%of one origami, keeping the more possible other characteristic constant. 
%figure presentation

\subsection{Understanding DNA Origami design}
In this section, we will rely on the model developed in the previous section  to explore how the melting temperature of
the rectangular origamis depends on the specific connectivity. This type of considerations could be relevant in applications
where it is necessary to improve
the stability of the template against temperature (DNA origamis  as platforms controlling chemical reactions or other applications
including grafting inorganic species).
\label{sec-2_1}
\subsubsection{Circular permutation of the scaffold.}
\label{sec-2_1_1}

\begin{figure}
\center
\includegraphics[width=0.9\textwidth]{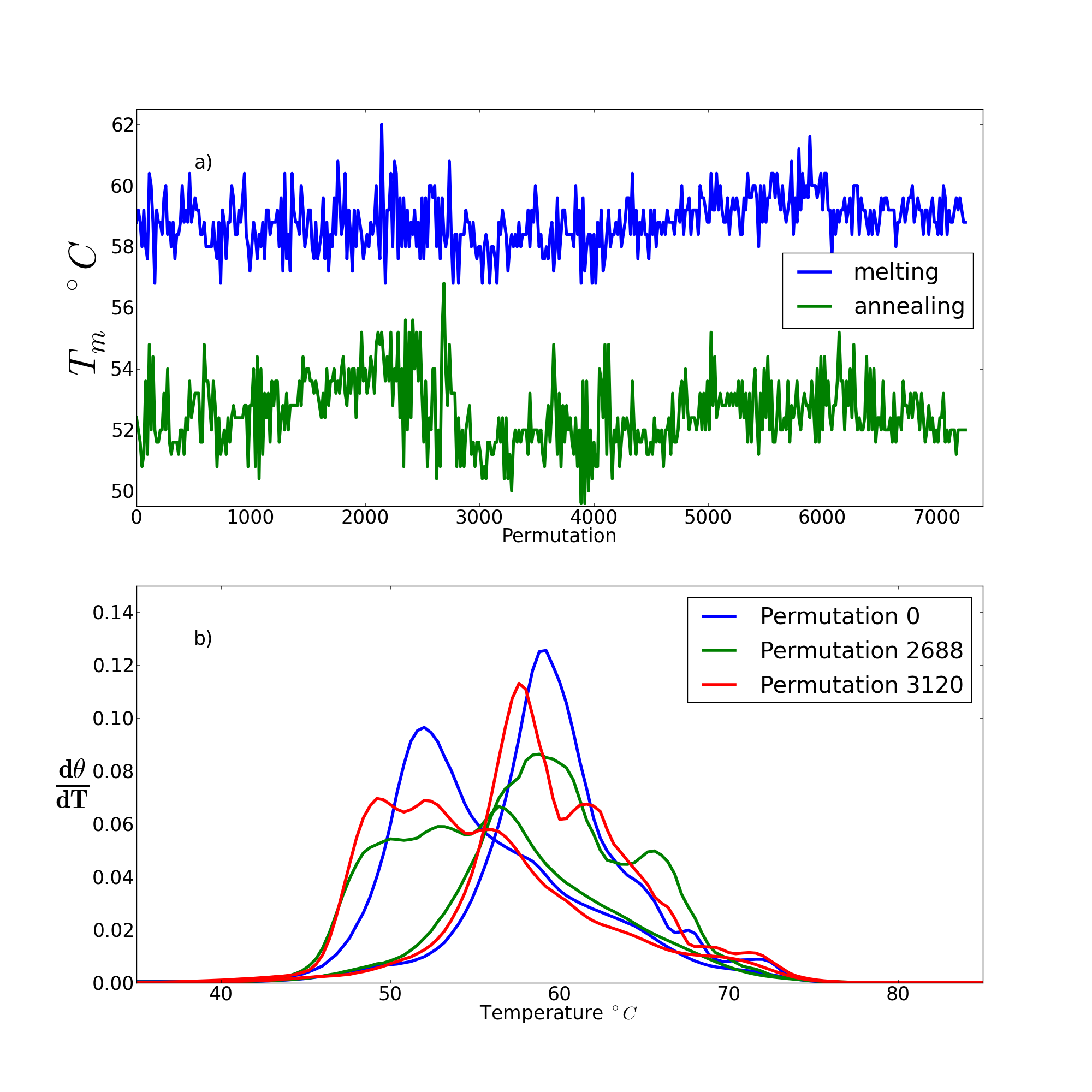}
\caption{(a) Distribution of melting temperatures (annealing and melting) as a function of the order of the circular permutation of the scaffold strand. (b)  two different melting curves corresponding to two permutations
with the lowest and highest annealing temperatures.}
\label{fig:permutation}
\end{figure}

The scaffold used for the design of rectangular origamis is a circular phage. Thus, it is possible to choose the beginning of the
scaffold sequence anywhere so that 7248 different sets of staples (with same length and position, but with a different composition) are possible.
We compared the melting curves given by those permutations on the O4 origami, permuting
the sequence in steps of 16 bases as this shifts the middle of one staple to the middle of the nearby staple.
The distribution of temperatures corresponding to the  maximum of the derivative for the annealing and melting curves (Fig.~\ref{fig:permutation}a)
shows an amplitude of variation of about 4 \dcelcius. 
%Also,   correlations exist between consecutive permutations.
Depending on the permutation, the melting curves can be very different in shape (Fig.~\ref{fig:permutation}b).
We conclude that, for a given scaffold, there is some sort of invariance regarding the choice of the
origin.

%compPerm.py difference

\subsubsection{Decreasing the number of crossovers.}
\label{sec-2_1_3}

\begin{figure}
\center
\includegraphics[width=0.9\textwidth]{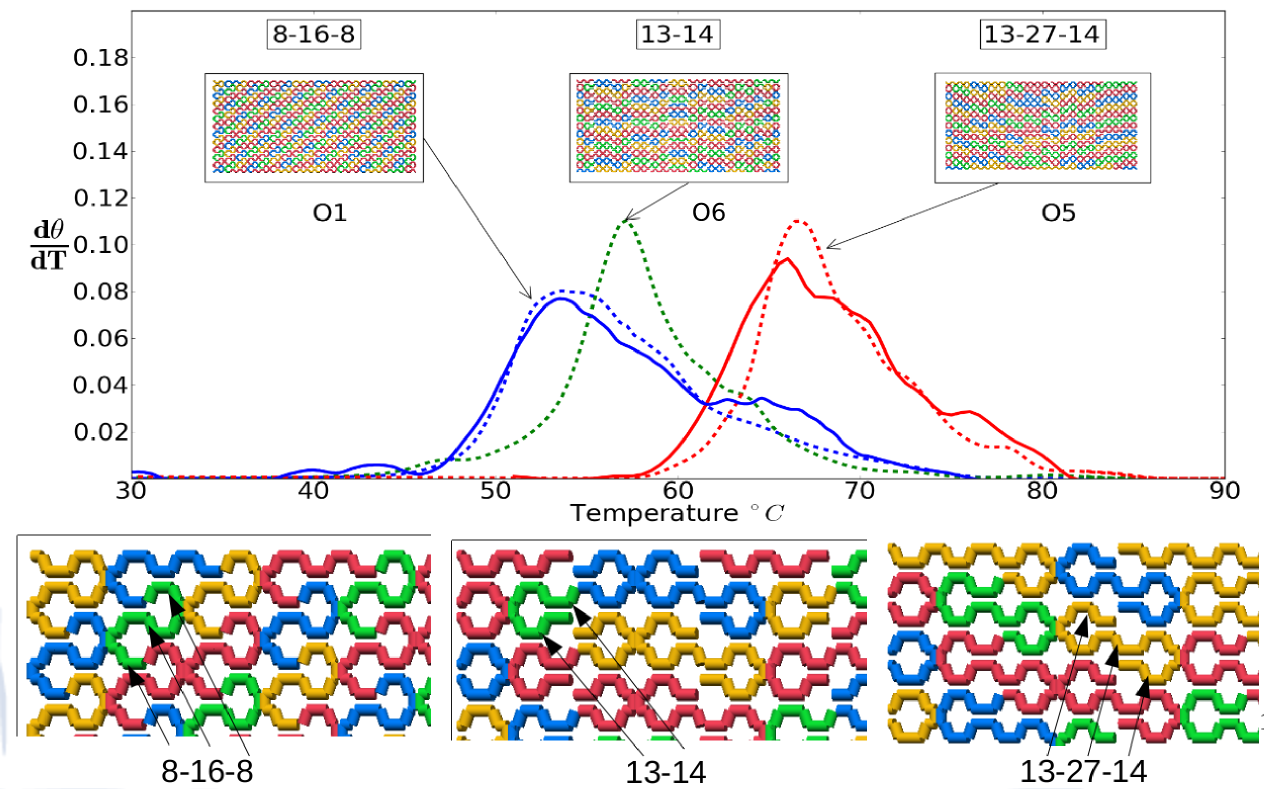}
\caption{Annealing curves of the O1, O5 and O6 origamis. The three origamis correspond to the same
scaffold pattern, but different staple pattern (solid line = experimental data, dashed line = theoretical curves). The
lower panel represents a detailed view of the differences in connectivity between the three origamis.}
\label{fig:dependence}
\end{figure}

In our model, a penalty is associated to each crossover. Reducing the number of crossovers should
in principle increase the stability in the annealing process. 
We started from the O1 shape to reduce the number of crossovers, as its design is more regular. 
In the initial origami there is a length of 32 bases between two crossovers, which corresponds to three periods in the double helix. Increasing the distance between crossovers leads to consider 54 bases (5 double-helix periods).
We considered two possibilities, illustrated in Fig.~\ref{fig:dependence}: staples  27b long, split in 13-14 (O6 origami), and staples
54b long, split in 13-27-14 (O5 origami). Indeed, there is a trade-off between the gain in enthalpy  when increasing the length of the staple, 
and the additional penalty of having two crossovers/ staple instead of only one. Our model shows that the net gain in stability,
compared to the initial 8-16-8 staple strategy, is almost 20 \dcelcius (Fig.~\ref{fig:dependence} O5). Again, the comparison with the experiments is excellent.
Notice that decreasing the number of crossovers could have an impact on the flexibility of the origami. 

\section{Conclusion}

\label{sec-3}
DNA origamis appear as a versatile tool to design various types of DNA based nanostructures.
We have introduced a simple algorithm based on known thermodynamic properties of dsDNA
and their parameterization with the NN-model \cite{santalucia2004thermodynamics}. The energy model
is based on a preliminary study on small origamis that points to the importance of crossovers and
their associated entropic penalty and correlation effects. 
This algorithm provides a reasonable account of the observed melting and annealing behaviour of DNA origamis.
  The model reproduces hysteresis and melting temperatures, as well as the width of
the melting curve. The hypothesis of local equilibrium turns out to be compatible
with the experimentally observed hysteresis. 
The model emphasizes the role of cooperativity in the folding process by introducing
correlations between the probability of presence of neighbour staples. This shows that the folding of each staple is strongly dependent
on the existence of a nearby cluster of folded staples. An intuitive picture can be derived from this modelling. In the annealing
process, only the staples with a high GC content are able to fold at high temperature. These staples provide with 
initial nuclei that  facilitate the folding of the neighbour staples. Therefore, the folding of staples with high AT content 
will take place at a temperature which is dependent on the distance to the set of these initial nuclei. During the melting process,
the unfolding of any staple is mostly independent on its neighborhood and it is well accounted by an average penalty in free energy.  
The histeresis is then explained by the fact that the
neighborhood of each staple is different in the annealing or melting.  

As a possible application, we have shown that the model allows to improve the
thermal stability by quantifying the effect of different construction factors such as staple length and density of
crossovers.  Extensions to 3D \cite{douglas2009self} \cite{dietz2009folding} and structures other than origami  \cite{hansen2010weave} are envisioned, as well as tests at the single molecule level (FRET).
As shown in \cite{oriAFM}, the interaction with the depositing surface will be crucial to interpret future AFM measurements on 
the folding of DNA origamis.

\begin{acknowledgments}
This work was partially supported by CNRS (PIR funding).
Thanks to Carmelo di Primo for letting us use its equipment.
Thanks to Jean-Louis Mergny for letting us use its equipment and to Thao Tran
for the Origami O2.
\end{acknowledgments}

\appendix

\section{Computing the probabilities from the law of mass action.}
\label{sec-4}

\label{sec:detail}

\begin{figure*}
\centering
\includegraphics[width=0.9\textwidth]{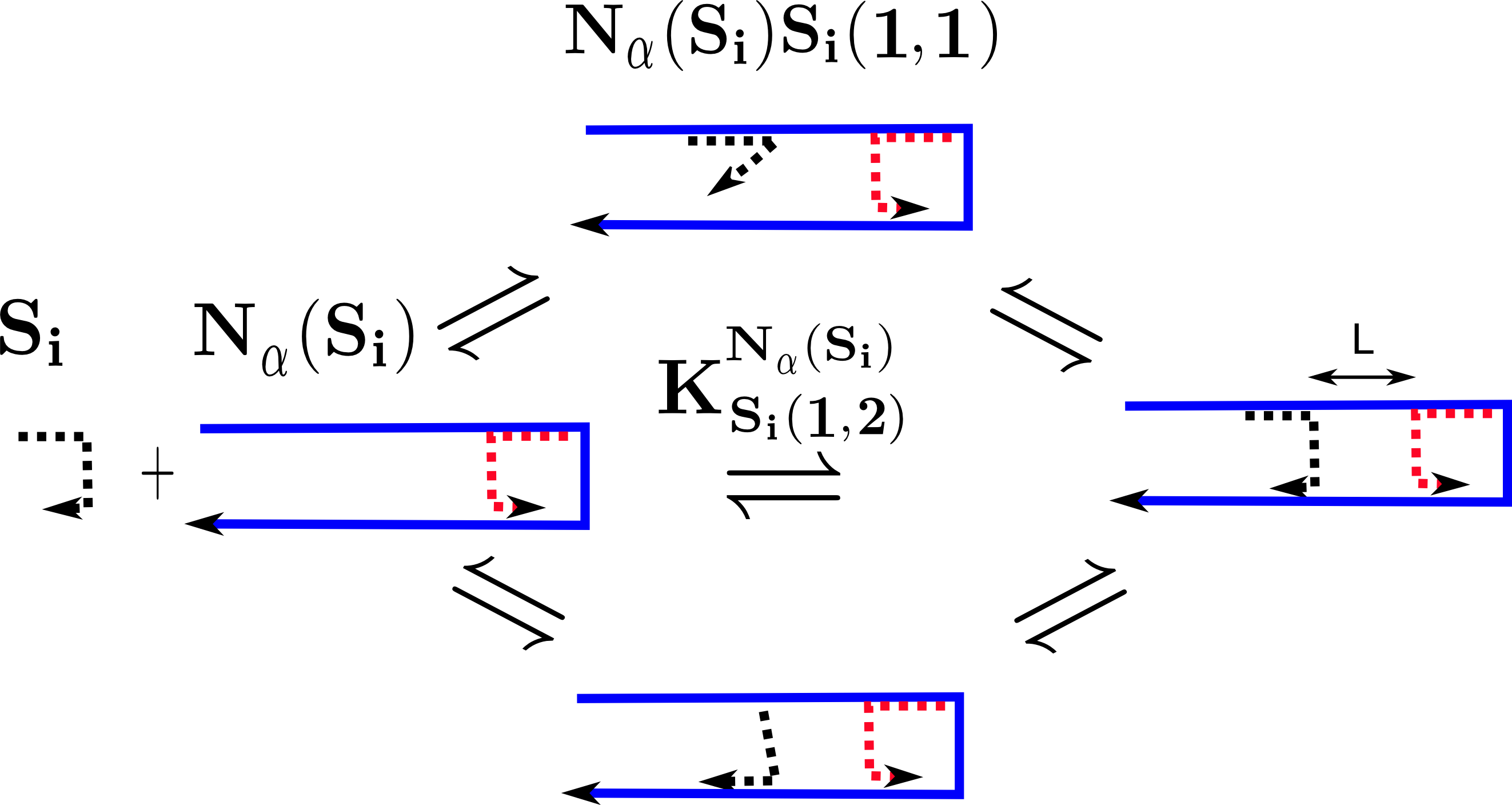}
\caption{Different possibilities for the staple $S_i$ to bind in the neighbourhood $ N_{\alpha}(S_i) $ }
\label{fig:stateM}
\end{figure*}

In this section, we explain how to compute the probability $p(S_i(k,l),T)$ by solving
a set of coupled equations which reflect the assumption that  the
reaction 
\[ S_i(k,l) + N_{\alpha}(S_i)\rightleftharpoons N_{\alpha}(S_i)S_i(k,l) \]
is governed by the law of mass action. For the sake of clarity,
 we consider first the particular case illustrated in Fig. \ref{fig:stateM}: a staple divided in
two parts of equal length is inserted in the vicinity of another staple that holds together 
a portion of the scaffold.  For this particular example, we nees to consider three equilibriums:
 the simultaneous binding of the two parts of $S_i$, with equilibrium constant:
\begin{equation}
K^{N_{\alpha}(S_i)}_{ S_i(1,2)} = \exp\left(-\frac{\Delta G_{NN}(S_i(1,2),T)}{kT}
\exp(-\Delta S_{bulge}(2L))/k \right)
\end{equation}
 The binding of only one half $S_i(1,1)$ of the staple, with equilibrium constant:
\begin{equation}
K^{N_{\alpha}(S_i)}_{ S_i(1,1)} = \exp\left(-\frac{\Delta G_{NN}(S_i(1,1),T)}{kT}  \right)
\end{equation} 

 The binding of the other half $S_i(2,2)$ of the staple:  
\begin{equation}
K^{N_{\alpha}(S_i)}_{ S_i(2,2)} = \exp \left(-\frac{\Delta G_{NN}(S_i(2,2),T)}{kT}  \right)
\end{equation}

More generally, for any set of equilibriums between $S_i$ and its neighborhood $N_{\alpha}(S_i)$, the
law of mass action reads:

\begin{equation}\label{eq:K}
\frac{[S_i(l,p)N_{\alpha}(S_i)]}{[N_{\alpha}(S_i)][S_i]} = K^{N_{\alpha}(S_i)}_{ S_i(l,p)} = exp\left(-\frac{\Delta G(S_i(l,p),T)}{kT} \right)
\end{equation}

The concentration of free staples in solution is given by
\[ C_0\left( Exs - \sum_{\beta}\sum_{u,v,u\le v} p(S_i(u,v)| N_{\beta}(S_i))p(N_{\beta}(S_i)) \right)\]
$Exs$ is the excess of staple concentration, compared to the initial concentration of scaffold $C_0$,
$p(S_i(u,v)| N_{\beta}(S_i))$ the conditional probability to observe $S_i(u,v)$ given the neighbour staples
$N_{\beta}(S_i)$ and $p(N_{\beta}(S_i))$ the probability of  the neighborhood $N_{\beta}$.
In order to get a closed set of equations, we approximate $p(N_{\beta}(S_i),T) \sim p(N_{\beta}(S_i),T-dT)$
where $dT$ is the temperature step in the melting-annealing process.
 
 As the excess in DNA origami annealing experiments is important, the concentration of free staples will vary weakly in the process of formation.
It is then approximated by $C_0 Exs$. 
%This aproximation is necessary to obtain a set of uncoupled equations.
The concentration of free $N_{\alpha}$ ($S_i$ not bounded) is given by:
\[ C_0\left( p(N_{\alpha}(S_i)) - \sum_{u,v,u\le v} p(S_i(u,v)| N_{\alpha}(S_i))p(N_{\alpha}(S_i)) \right).\]

The concentration of the configuration $S_i(l,p)$ hybridized in the neighborhood $N_{\alpha}(S_i)$ is
\[  C_0 p(S_i(l,p)| N_{\alpha}(S_i))p(N_{\alpha}(S_i)).  \]

The approximated expression of the equilibrium constant is therefore:
\begin{eqnarray}
\lefteqn{K^{N_{\alpha}(S_i)}_{ S_i(l,p)} } \nonumber \\  & \sim & 
%\frac{MS_i(l,p)}{S_iC_{N_{\alpha}(S_i)}}= \\
\frac{ p(S_i(l,p) | N_{\alpha}(S_i))p(N_{\alpha}(S_i))}{  Exs C_0 \left( p(N_{\alpha}(S_i)) - \displaystyle{\sum_{u,v,u\le v}} p(S_i(u,v)| N_{\alpha}(S_i))p(N_{\alpha}(S_i)) \right)} \label{eqKe}
\end{eqnarray}

This set of equations can be solved by noticing that the denominator in \ref{eqKe} is independent of $S_i(l,p)$. Thus,
 for any couple of configurations $(S_i(l,p),S_i(u,v))$, 
\begin{equation}\label{eqa:frac}
 \frac{p(S_i(l,p)|N_{\alpha}(S_i))}{p(S_i(u,v)|N_{\alpha}(S_i))} = \frac{K^{N_{\alpha}(S_i)}_{ S_i(l,p)}}{K^{N_{\alpha}(S_i)}_{ S_i(u,v)}}
\end{equation}
Let $y_{\alpha} = \displaystyle{\sum_{l,p,l \le p}} p( S_i(l,p)|N_{\alpha}(S_i))$ and  $S_{\alpha} = \displaystyle{\sum_{l,p,l \le p}} K^{N_{\alpha}(S_i)}_{ S_i(l,p)}$. Then, by summing over $l$ and $p$ eq. \ref{eqKe} reduces to:

\begin{equation}
\frac{1}{C_0 Exs}\frac{ y_{\alpha} }{1-y_{\alpha} } = S_{\alpha}
\end{equation}
 
or:

\begin{equation}
y_{\alpha} = \frac{C_0 Exs S_{\alpha}}{1+C_0 Exs S_{\alpha}}
\end{equation}

and finally
\begin{equation}
p(S_i(l,p),T) =\displaystyle{ \sum_{\alpha}}p(S_i(l,p)|N_{\alpha})p(N_{\alpha}(S_i),T) \sim 
\displaystyle{ \sum_{\alpha}} p(N_{\alpha}(S_i),T-dT)y_{\alpha} \frac{ K^{N_{\alpha}(S_i)}_{ S_i(l,p)}}{S_{\alpha}}
\end{equation}

\section{Temperature and salt corrections}
\label{TScorr}
The computation of the nearest-neighbour contribution $\Delta G_{NN}= \Delta H_{NN} - T \Delta S_{NN}$ includes corrections to take
into account temperature and salt variations.
\subsection{Temperature corrections}

Besides the nearest-neighbour contributions ($\Delta H_{N}^0,\Delta S_{N}^0$) \cite{santalucia2004thermodynamics}, we have
also included temperature dependent corrections:
\begin{eqnarray*}
\Delta H_{NN}(S_i(l,p))& = &\Delta H_{N}^0(S_i(l,p)) + C_p|l-p+1|(T-T_{ref}) \\[6pt]
\Delta S_{NN}(S_i(l,p))& = &\Delta S_{N}^0(S_i(l,p)) + C_p|l-p+1| \ln (\frac{T}{T_{ref}})
\end{eqnarray*}

where
 $C_p = -42 cal/mol/K/bases$ and $T_{ref} = 53 ^{\circ}C$ according
 to \cite{hughesman2011correcting}.

\subsection{Salt corrections}

The parameters of the NN model \cite{santalucia2004thermodynamics} are given for standard salt concentrations ([Na]=1M, [Mg]=0).
Different salt conditions can be taken into account using the correcting terms in  \cite{owczarzy2008predicting}. These corrections apply when Mg is dominant.
We assume that for each attachment of the staple $S_i$ on the configuration $S_i(l,p)$ (of length $|S_i(l,p)|$),
the relation
\begin{equation*}
\frac{1}{T_m(Mg,Na,C,S_i(l,p))} = \frac{1}{T_m(0,Na,C,S_i(l,p))} + f(Mg,Na,f_{GC},|S_i(l,p)|)
\end{equation*}
between the melting temperatures $T_m$ at different concentrations of Mg and Na holds ($f_{GC}$ is the fraction of GC content of $S_i(k,l)$).

The melting temperature corresponding to the point where $\Delta G (T_m)= 0$:

\begin{equation}
T_m(0,Na,C,S_i(l,p)) = \frac{\Delta H_{NN}(S_i(l,p) )  }{ \Delta S_{NN}(S_i(l,p)) + R ln(C/4)}
\label{eq:Tm}
\end{equation} it can be deduced that the salt corrections are taken 
into account by an entropic correction  given by:
\begin{equation}
SC(Mg,Na) = f(Mg,Na,f_{GC},|S_i(l,p)|) \Delta H
\end{equation}
where $\Delta H$ is the sum of all the contributions previously cited 
(nearest-neighbour, temperature corrections, topological contributions).
 In this paper, we modified the function $f(Mg,Na,f_{GC},|S_i(l,p)|)$ as
given in \cite{owczarzy2008predicting}  by a small additive term to take into account low AT content strands.
In order to do this, we compared the predictions of \cite{owczarzy2008predicting}  with the experimental results obtained with the
small origamis. In \cite{owczarzy2008predicting}, the authors calibrated their model against 17 different dsDNA, involving a wide
range of salt concentrations, 
and obtained a mean deviation  $-1.7  \pm  0.7$ \dcelcius. We deal here with a more restricted range of salt concentrations, it is
therefore expected that the model \cite{owczarzy2008predicting} can be improved. We used a correction to the $f$ function of 
\cite{owczarzy2008predicting} for the low fractions $f_{GC}$ of GC content, 
to obtain a mean deviation of $-0.4 \pm 0.3$ \dcelcius. The corrected $f$ function reads:

\[f_{new} = f + (f_{gc}-0.5) 0.00008 \; \; \; {\rm if} \; \; \; f_{gc}<0.5.\]

\begin{figure}
\center
\includegraphics[width=0.6\textwidth]{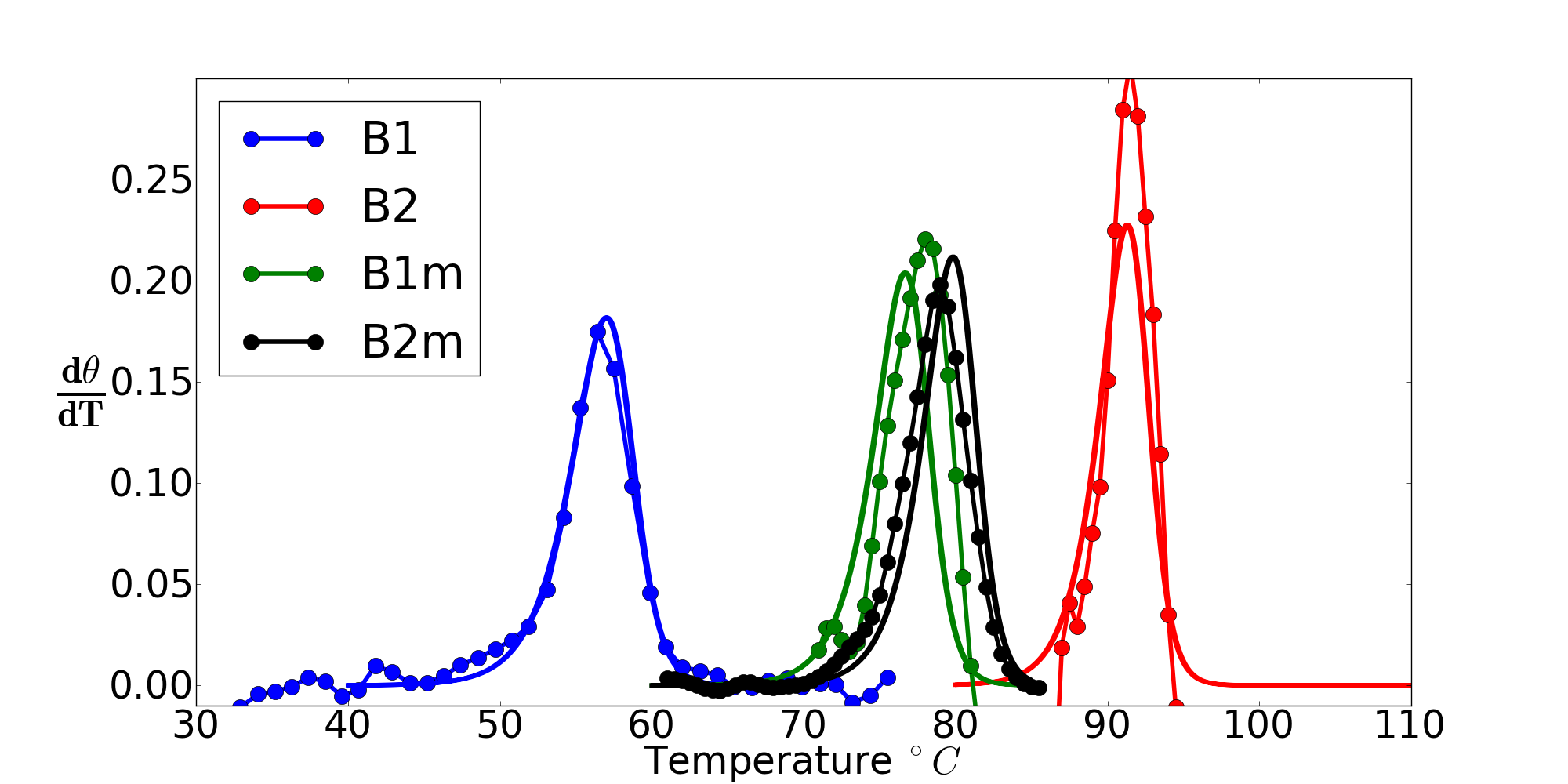}
\caption{(dotted line = experiment, solid line = theoretical) Comparison between the model and the melting experiments for four
different dsDNA}
\label{fig:brinAlone}
\end{figure}

The results obtained with this new expression for the strands involved in the differents small origamis
are illustrated in  Fig. \ref{fig:brinAlone}.

\newpage

\end{document}